\documentclass[11pt,twoside]{article}
\usepackage{./asp2014}
\usepackage{color}
\usepackage{gensymb}
\aspSuppressVolSlug
\resetcounters

\newcommand{\Lsun}{\ensuremath{\rm L_\odot}}

\begin{document}

\title{On the nature of  anomalous reddening of Cygnus~OB2~\#12 hypergiant}
\author{O.\,V.\,Maryeva$^1$, E.\,L.\,Chentsov$^1$, V.\,P.\,Goranskij$^2$, V.\,V.\,Dyachenko$^1$ and S.\,V.\,Karpov$^{1,3}$
\affil{$^1$Special Astrophysical Observatory of the Russian Academy of Sciences, Nizhnii Arkhyz, 369167, Russia; \email{olga.maryeva@gmail.com}}
\affil{$^2$Sternberg Astronomical Institute, Moscow State University, Universitetsky pr., 13, Moscow, 119992, Russia}
\affil{$^3$Astronomy and  geodesy department of Kazan (Volga region) Federal University,Kremlevskaya str., 18, Kazan, 420008, Russia}}


\paperauthor{O.\,V.\,Maryeva}{olga.maryeva@gmail.com}{ORCID_Or_Blank}{Special Astrophysical Observatory of the Russian Academy of Sciences}{Astrospectroscopy Laboratory}{Nizhnii Arkhyz}{Karachai-Cherkessian Republic}{369167}{Russia}
\paperauthor{E.\,L.\,Chentsov}{echen@sao.ru}{ORCID_Or_Blank}{Special Astrophysical Observatory of the Russian Academy of Sciences}{Astrospectroscopy Laboratory}{Nizhnii Arkhyz}{Karachai-Cherkessian Republic}{369167}{Russia}
\paperauthor{V.\,P.\,Goranskij}{??????}{ORCID_Or_Blank}{}{Sternberg Astronomical Institute }{Moscow}{ }{119992}{Russia}
\paperauthor{V.\,V.\,Dyachenko}{?????}{ORCID_Or_Blank}{Special Astrophysical Observatory of the Russian Academy of Sciences}{Astrospectroscopy Laboratory}{Nizhnii Arkhyz}{Karachai-Cherkessian Republic}{369167}{Russia}
\paperauthor{S.\,V.\,Karpov}{????}{ORCID_Or_Blank}{Special Astrophysical Observatory of the Russian Academy of Sciences}{Astrospectroscopy Laboratory}{Nizhnii Arkhyz}{Karachai-Cherkessian Republic}{369167}{Russia}

\begin{abstract}
 To explain the nature of the anomalously high reddening ($A_V\simeq 10$~mag) towards one of the most luminous stars in the Galaxy -- Cyg~OB2~\#12 (B5 Ia-0), also known as MT304,  we carried out spectro-photometric observations of 24 stars located in its vicinity.   We included  five new B-stars  to the members of Cygnus OB2, and for five more photometrically selected  stars  we spectroscopically confirmed their membership.  We constructed the map of interstellar extinction within 2.5~arcmin radius  and found that interstellar extinction increases towards  MT304. The increase of reddening suggests that the reddening excess may be caused by the circumstellar shell ejected by the star during its evolution. We also report the  detection of  a second component of MT304, and discovery of  an even fainter third component, based  on data of speckle  interferometric observations taken with the Russian 6-m telescope.
\end{abstract}

\subsection*{Introduction}

      Stellar association Cyg~OB2 (VI~Cyg) is one of the closest star formation regions to the Sun (distance is  $1.4\pm0.1$~kpc \citep{Rygl}) and it is a zoo of unique objects among which there is an {\it enigmatic} \ star Cyg~OB2~\#12 (Schulte~12, MT304). MT304\footnote{MT is the catalogue of Cyg~OB2 members by \citet{MT91}} is  {\it enigmatic} not only due to its high luminosity (its bolometric luminosity is $1.9\cdot 10^6 \Lsun$ \citep{Clark} and according to various estimates, the star is one of the brightest stars in the Galaxy \citep{Clark}), but also because of its strong reddening, which is higher than average reddening of the massive stars in the association (see \citet{Chentsov} and references therein, and \citet{Wright2015}). 
   
      MT304 is classified as  B5 Ia-0 \citep{Chentsov} or B3-4 Ia \citep{Clark} hypergiant. Compared to other bright massive stars in the association  MT304 is significantly more reddened. Interstellar extinction towards it is  $A_V\simeq10.1$~mag \citep{KiminkiAv,Wright2015}. According to \citet{Wright2015} the difference between MT304 and MT488, second most reddened OB-star, is $\Delta A_V=1.9$~mag. Nature of this excessive absorption remains unclear. Does it originate in a small dense dust cloud which is accidentally caught in the line of sight (see \citet{Whittet2015} and discussion therein)? Or does it arise in a circumstellar shell (see for example \citet{Chentsov})? 
  
      In order to investigate the extinction near MT304 we carried out long-slit spectroscopy and photometry of stars  within  2.5~arcmin  from it. Moreover we refined parameters of the second companion of MT304 using speckle-interferometer observations with the Russian 6-meter telescope. In this article we will report the results of these observations. 
      
\subsection*{Multiplicity of MT304}\label{sec:res}

  \citet{Caballero} discovered a second component of MT304, with the separation between the components of $63.6~{\rm mas}$ and the magnitude difference of $\Delta $m=$2.3\pm0.2$~mag in F583W filter. To confirm it, and to measure the brightness difference in other spectral bands, as well as to  better study the immediate neighbourhood of MT304 in general, we performed speckle-interferometric observations with the Russian 6-m telescope on February 12 and December 5 2014. Observations were carried out in the visual and near-infrared spectral ranges (Table~\ref{tab:specl}).

\vspace{-5mm}
\begin{table*}
\caption{Resolved companions of MT304. $\theta$ is the measured position angle, $\rho$ is the measured angular separation, $\Delta$m is the observed magnitude difference, $\lambda$ is the central wavelength of the filter used for the observation, $\Delta\lambda$ -- the full width at half-maximum (FWHM) of the filter passband. $^*$ Data taken from \citet{Caballero}}
\label{tab:specl}
\begin{tabular}{lcccccc}
\hline
      Pair   &   Epoch   & $\theta$, [$\degree$]       & $\rho$         & $\Delta$m      & $\lambda$ & $\Delta\lambda$ \\
             &           &                             & [mas]          & [mag]          & [nm]      &  [nm]          \\
\hline
AB           & 2014.1198 & $293.0\pm 0.3$              & $65\pm   1$    & $1.79\pm 0.02$ & 700       & 40             \\
             & 2014.1199 & $293.0\pm 0.3$              & $64\pm   2$    & $1.75\pm 0.03$ & 800       & 100            \\
             & 2014.9290 & $291.7\pm 0.3$              & $64\pm   2$    & $2.0 \pm 0.1 $ & 800       & 100            \\
             & 2014.9290 & $291.7\pm 0.3$              & $69\pm   2$    & $1.8 \pm 0.1 $ & 900       & 80             \\
\\
AC           & 2014.1198 & $271.3\pm 0.2$              & $1246\pm 2$    & $4.8 \pm 0.2 $ & 700       & 40             \\
\hline
AB$^*$       & 2006.2346 & ${305.9\atop283.5} \pm 3.3$ & $63.5\pm3.5$   & $2.3\pm0.2$  & 583       & 234            \\ 
\hline
\end{tabular}
\end{table*}
\vspace{-3mm} 

\begin{figure*}\begin{center}
 \includegraphics[scale=0.248,viewport=80 28 678 655,clip]{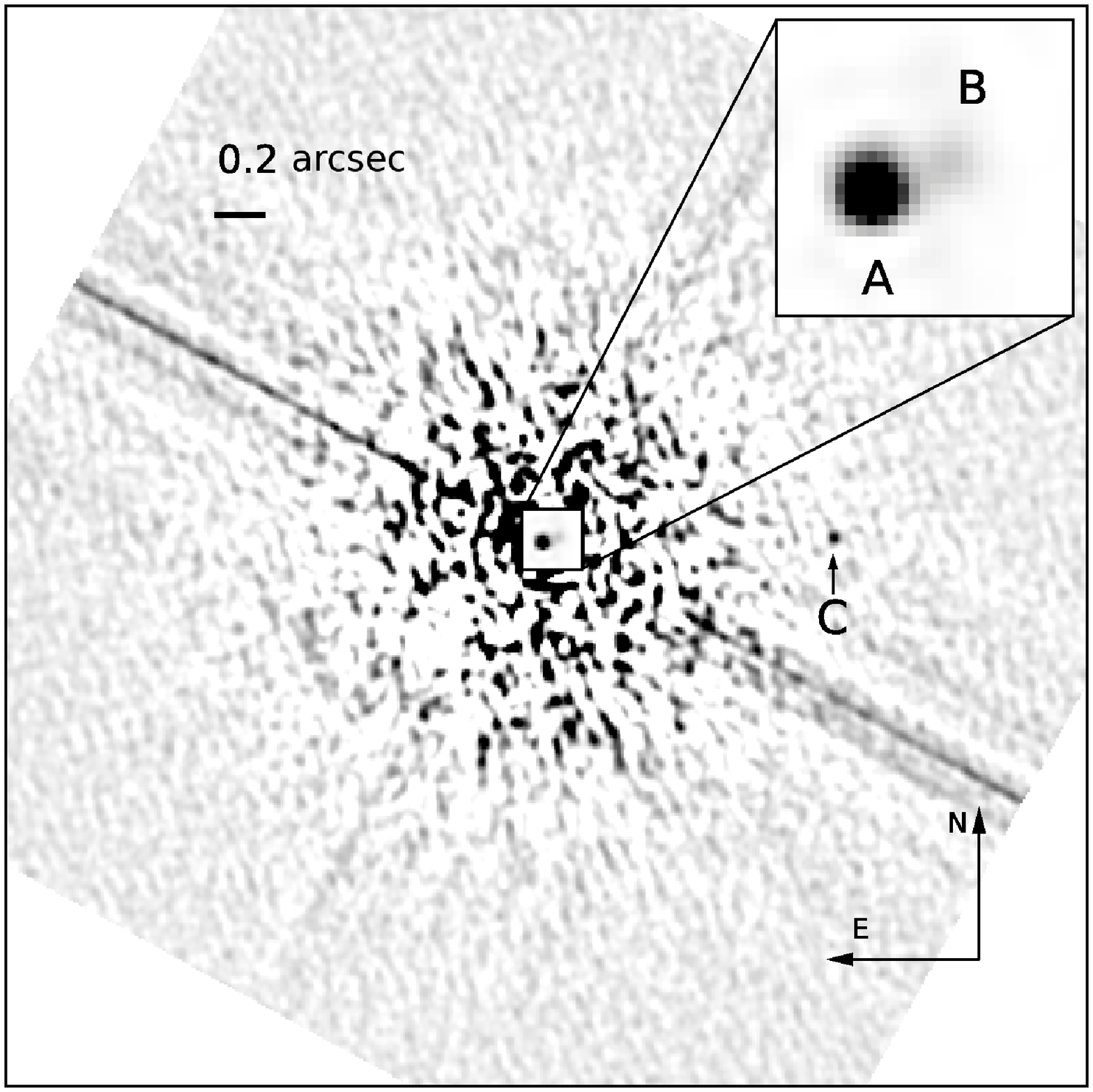}
\includegraphics[scale=0.45]{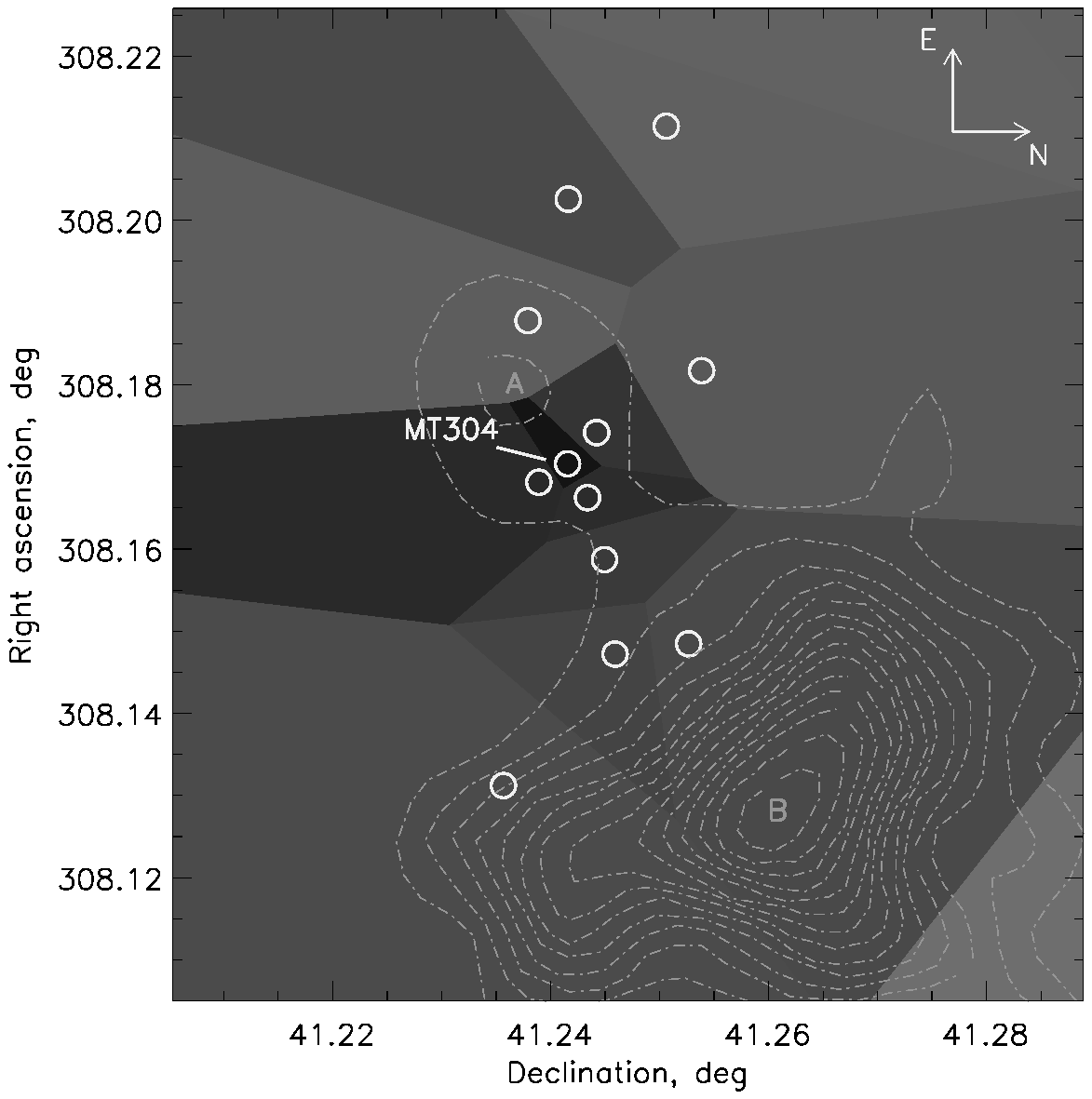}
\end{center}
\caption{Left: the autocorrelation functions of MT304 multiple system in the 9000/800~\AA \ filter. Right:  Voronoi tessellated extinction map (one star per Voronoi cell) for twelve sample stars (except for USNO-B1.0 1312-0389914) belonging to Cyg~OB2. Dash-dotted lines  show contours of the $^{13}$~CO integrated intensity $\int T_A^*~dv$.}
\label{fig:map2}
\end{figure*}

   The autocorrelation function (see left panels of Figure~\ref{fig:map2}) clearly displays the second component of MT304, discovered by \citet{Caballero}. Moreover, the fainter third companion is also seen with the significance better than at least 7 sigmas. 
   The measured parameters are given in Table~\ref{tab:specl}; for the second component they are well consistent with the estimates by \citet{Caballero}.
   
   We conducted speckle interferometric observations of MT304 in February and December 2014. As can be seen in Table~\ref{tab:specl} during these 10 months the position angle of the secondary component  has changed by 1.7~$\deg$, greatly exceeding the measurement errors. It strongly favours the physical connection between A and B components. Our measurements together with the results of \citet{Caballero} suggest a rotation period of the secondary component of about 100-200 years. The motion of the secondary component is large enough to build the orbit of the system in ten years of observations and therefore to derive the mass ratio of these stars.
  
   The observed spectrum of MT304 does not display any lines of the secondary component (see e.g. \citet{Chentsov}) and is perfectly fit by a single star model \citep{Clark}, and therefore we may suggest that the secondary component is a B dwarf with spectrum similar to the one of  the brighter star. Dedicated observations with high signal to noise ratio are definitely necessary to classify it better. We can not presently say anything about the third component, but its spatial proximity suggests that it may belong to the Cyg OB2 association and be physically connected to MT304 too. We will perform further observations of this system. 

 \subsection*{Reddening around MT304}\label{sec:obs}  
 
      Longslit spectroscopy of stars close to MT304 was carried out with the Russian 6-m telescope with the focal reducer  SCORPIO \citep{scorpio}, using the grism VPHG~1200G (spectral resolution $\Delta\lambda\simeq6.5$, spectral range 4000--5700~\AA). All the SCORPIO spectra were reduced using the {\tt ScoRe} package\footnote{{\tt ScoRe} package \url{http://www.sao.ru/hq/ssl/maryeva/score.html}}. 
      
      Images of the region around MT304 in {\it B} and {\it V} filters have been acquired using the direct imaging mode of SCORPIO focal reducer and  the CCD photometer of Zeiss-1000 telescope of SAO RAS.

   We performed the spectral classification of the stars in our sample. By combining estimates of spectral types with the photometric data we measured the individual interstellar extinction ${\rm A}_{V}$. We analyzed the stars with V=13-20~mag and we increased the number of highly reddened massive stars, reducing the difference in reddening between MT304 and other members of the association. Before our study MT488 was the second reddened massive star ($\Delta A_V=1.8$~mag) after MT304. Now J203240.35+411420.1 and J203239.90+411436.2 are the most reddened  massive stars and the difference is reduced down to $0.9\pm0.1$~mag. However MT304 remains the most reddened among massive stars in Cyg~OB2.
                  
\begin{figure*}\centering
\includegraphics[scale=0.3,viewport=40 20 530 530,clip]{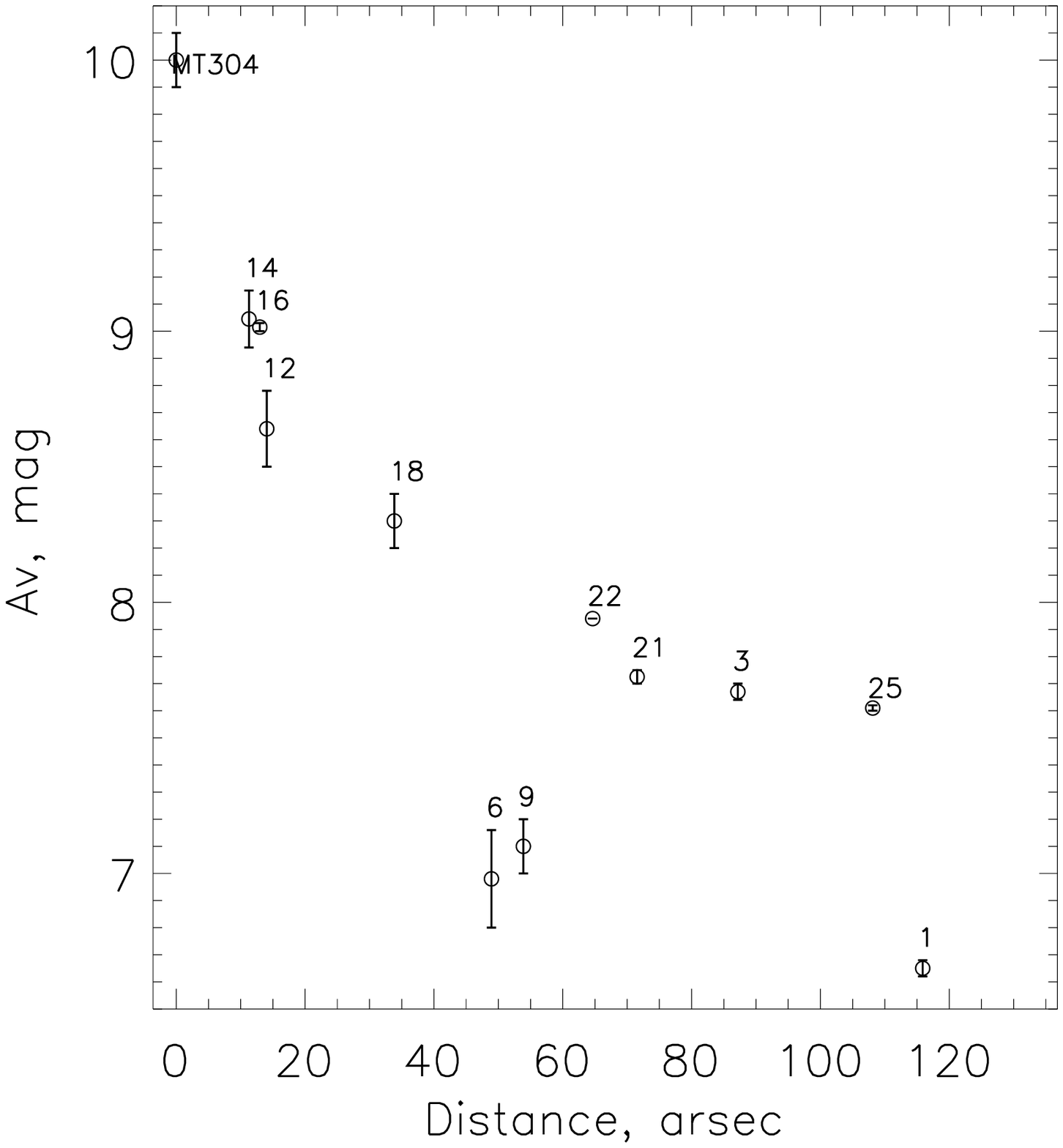}
\includegraphics[scale=0.3,viewport=40 20 530 530,clip]{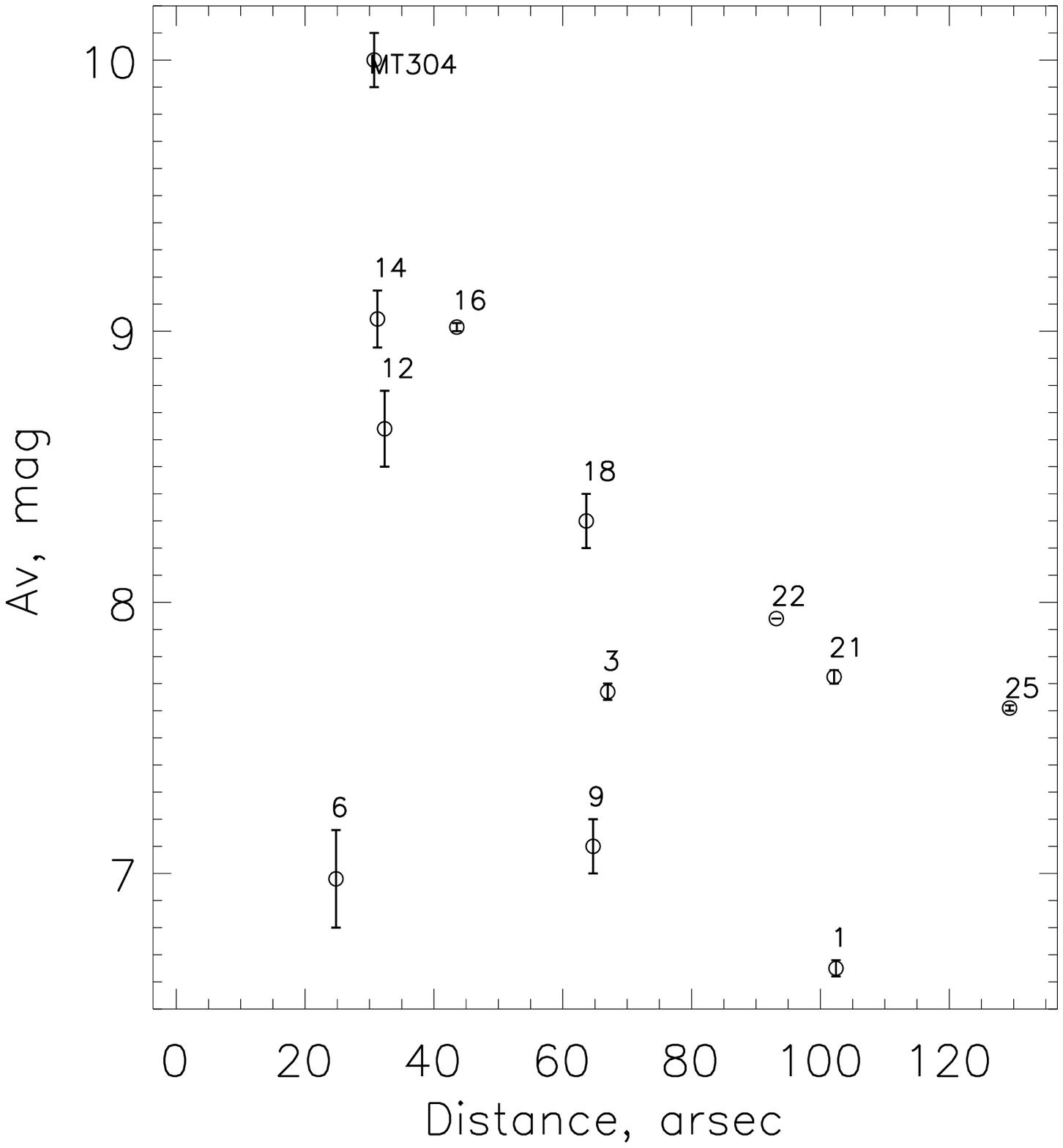}
\caption{Dependency of the interstellar extinction  $A_V$ on the distance to MT304 (left part) and on the distance to core A of the clump (right part).}
\label{fig:distav}
\end{figure*}

   Figure~\ref{fig:map2} shows the map of interstellar extinction based on the obtained data. Extinction clearly increases with approaching to MT304, as it is shown in left panel of Figure~\ref{fig:distav}. For the stars located within 30~arcsec from MT304 the extinction is higher than 8.5 mag.

 \subsection*{Conclusions}\label{sec:concl}

    To study the high reddening affecting the hypergiant Cyg~OB2~\#12 (MT304) we conducted longslit spectroscopy and photometry of 24 stars with V=13-20~mag laying within 2.5 arcmin from this hypergiant. For 22 of these stars the spectroscopy was performed for the first time. Spectral analysis shows  that 13 of the studied stars, along with MT304, belong to Cyg~OB2. Thus, the extended the list of Cyg~OB2 massive stars by ten more B stars. 
    
    Interstellar extinction increases while approaching to MT304. J203240.35+411420.1 and J203239.90+411436.2 are the most reddened massive stars after  MT304. They are located  within 13 and 15~arcsec away from MT304 and their $A_V$ are $9.1\pm0.1$~mag and $9.02\pm0.02$~mag respectively.

    Our speckle-interferometer observations confirmed  that MT304 has a second companion, which was discovered by \citet{Caballero}. We detected its orbital motion which suggests that the orbital period of the system ranges from 100 to 200 years. We also observed for the first time a third component which is weaker than main component by 4.8 mag. 


 More details are published in the article by  \citet{me2016Balt} and \citet{MaryevaChentsov2016MNRAS}.       

\acknowledgements The study was supported by the RFBR (projects no. 16-02-00148, 14-02-00291, 14-02-00759). The work is performed according to the Russian Government Program of  Competitive Growth of Kazan Federal University. Sergey Karpov thanks the grant of Russian Science Foundation No~14-50-00043. 

\bibliography{Maryevabib}

\begin{thebibliography}{}
\expandafter\ifx\csname natexlab\endcsname\relax\def\natexlab#1{#1}\fi
\expandafter\ifx\csname url\endcsname\relax
  \def\url#1{\texttt{#1}}\fi
\expandafter\ifx\csname urlprefix\endcsname\relax\def\urlprefix{URL }\fi
\providecommand{\eprint}[2][]{\url{#2}}

\bibitem[{{Afanasiev} \& {Moiseev}(2005)}]{scorpio}
{Afanasiev}, V.~L., \& {Moiseev}, A.~V. 2005, Astronomy Letters, 31, 194

\bibitem[{{Caballero-Nieves} et~al.(2014){Caballero-Nieves}, {Nelan}, {Gies},
  {Wallace}, {DeGioia-Eastwood}, {Herrero}, {Jao}, {Mason}, {Massey}, {Moffat},
  \& {Walborn}}]{Caballero}
{Caballero-Nieves}, S.~M., {Nelan}, E.~P., {Gies}, D.~R., {Wallace}, D.~J.,
  {DeGioia-Eastwood}, K., {Herrero}, A., {Jao}, W.-C., {Mason}, B.~D.,
  {Massey}, P., {Moffat}, A.~F.~J., \& {Walborn}, N.~R. 2014, \aj, 147, 40

\bibitem[{{Chentsov} et~al.(2013){Chentsov}, {Klochkova}, {Panchuk}, {Yushkin},
  \& {Nasonov}}]{Chentsov}
{Chentsov}, E.~L., {Klochkova}, V.~G., {Panchuk}, V.~E., {Yushkin}, M.~V., \&
  {Nasonov}, D.~S. 2013, Astronomy Reports, 57, 527

\bibitem[{{Clark} et~al.(2012){Clark}, {Najarro}, {Negueruela}, {Ritchie},
  {Urbaneja}, \& {Howarth}}]{Clark}
{Clark}, J.~S., {Najarro}, F., {Negueruela}, I., {Ritchie}, B.~W., {Urbaneja},
  M.~A., \& {Howarth}, I.~D. 2012, \aap, 541, A145

\bibitem[{{Kiminki} et~al.(2007){Kiminki}, {Kobulnicky}, {Kinemuchi}, {Irwin},
  {Fryer}, {Berrington}, {Uzpen}, {Monson}, {Pierce}, \& {Woosley}}]{KiminkiAv}
{Kiminki}, D.~C., {Kobulnicky}, H.~A., {Kinemuchi}, K., {Irwin}, J.~S.,
  {Fryer}, C.~L., {Berrington}, R.~C., {Uzpen}, B., {Monson}, A.~J., {Pierce},
  M.~J., \& {Woosley}, S.~E. 2007, \apj, 664, 1102

\bibitem[{{Maryeva} et~al.(2016{\natexlab{a}}){Maryeva}, {Chentsov},
  {Goranskij}, {Dyachenko}, {Karpov}, {Malogolovets}, \&
  {Rastegaev}}]{MaryevaChentsov2016MNRAS}
{Maryeva}, O.~V., {Chentsov}, E.~L., {Goranskij}, V.~P., {Dyachenko}, V.~V.,
  {Karpov}, S.~V., {Malogolovets}, E.~V., \& {Rastegaev}, D.~A.
  2016{\natexlab{a}}, \mnras, 458, 491

\bibitem[{{Maryeva} et~al.(2016{\natexlab{b}}){Maryeva}, {Chentsov},
  {Goranskij}, \& {Karpov}}]{me2016Balt}
{Maryeva}, O.~V., {Chentsov}, E.~L., {Goranskij}, V.~P., \& {Karpov}, S.~V.
  2016{\natexlab{b}}, Baltic Astronomy, 25, 42

\bibitem[{{Massey} \& {Thompson}(1991)}]{MT91}
{Massey}, P., \& {Thompson}, A.~B. 1991, \apj, 101, 1408

\bibitem[{{Rygl} et~al.(2012){Rygl}, {Brunthaler}, {Sanna}, {Menten}, {Reid},
  {van Langevelde}, {Honma}, {Torstensson}, \& {Fujisawa}}]{Rygl}
{Rygl}, K.~L.~J., {Brunthaler}, A., {Sanna}, A., {Menten}, K.~M., {Reid},
  M.~J., {van Langevelde}, H.~J., {Honma}, M., {Torstensson}, K.~J.~E., \&
  {Fujisawa}, K. 2012, \aap, 539, A79

\bibitem[{{Whittet}(2015)}]{Whittet2015}
{Whittet}, D.~C.~B. 2015, \apj, 811, 110

\bibitem[{{Wright} et~al.(2015){Wright}, {Drew}, \& {Mohr-Smith}}]{Wright2015}
{Wright}, N.~J., {Drew}, J.~E., \& {Mohr-Smith}, M. 2015, \mnras, 449, 741

\end{thebibliography}


\end{document}